\documentclass[10pt]{article}
\usepackage{amets,amssymb,amsmath,amstext,lineno,color,xcolor,comment,grffile,array,multirow,makecell,nicefrac}
\usepackage[percent]{overpic}
\usepackage{hyperref,graphicx,multicol,pdfpages}
\hypersetup{colorlinks = true, allcolors = blue, allbordercolors = white}
\usepackage[round,authoryear]{natbib}
\usepackage{float,rotating,epic,tabularx}
\bibpunct[,]{(}{)}{;}{a}{}{,}
\def\negicm{-0.25cm}

\begin{document}
\renewcommand{\baselinestretch}{1.0}\small\normalsize

\small
\noindent
The following is an author-produced version of an article accepted for
publication in the journal Remote Sensing of Environment on January 27, 2023
(\textcopyright 2023 under the
\href{https://creativecommons.org/licenses/by-nc-nd/4.0}{CC-BY-NC-ND
  4.0 license}).  Initial submission was on March 14, 2022.
The definitive version from the publisher can be cited as
\href{https://doi.org/10.1016/j.rse.202x.xx.xxx}{Danielson, R. E.,
  H. Shen, J. Tao, and W. Perrie, 2023: Dependence of ocean surface
  filaments on wind speed: An observational study of North Atlantic
  right whale habitat.}  All versions seek to address:
% Remote Sens. Environ., xxx, xx-xx,
% doi:10.1016/j.rse.202x.xx.xxx.}  All versions seek to address:
\vspace{0.3in}

\noindent
Collocations of Gulf of St.~Lawrence whale sightings and synthetic aperture radar scenes\\
Quantification of a dependence of submesoscale filaments at the ocean surface on wind speed\\
Consistency in linear and nonlinear components of Pearson and distance correlation\\
A simple wind speed adjustment of synthetic aperture radar contrast\\

\begin{center}
\vspace{0.2in}
{\Large Dependence of ocean surface filaments on\\
wind speed: An observational study of\\
North Atlantic right whale habitat\\}
\vspace{0.3in}

\small
\noindent
Richard E. Danielson, Hui Shen, Jing Tao, and William Perrie\\
Bedford Institute of Oceanography, Fisheries and Oceans Canada\\ Dartmouth, Nova Scotia, Canada\\
\end{center}
\vspace{0.1in}

\begin{center}
\vspace{0.3in}

\large
Abstract\\
\end{center}
\normalsize

Coherent filaments at the ocean surface often appear to be transient
watermass boundaries, where currents converge, surfactants accumulate,
and frontal structure at depth can possibly delineate enhanced
biological activity in the upper water column.  Spaceborne synthetic
aperture radar (SAR) permits filaments to be observed at O[1-km]
resolution, but extensive coherent structures are more apparent in
weaker winds.  A wind speed adjustment is proposed for filaments
(i.e., contiguous SAR contrasts) of at least 10~km in length.
Measures of dependence (distance correlation and the linear and
nonlinear components of Pearson correlation) are examined to identify
a broad peak in the relationship between filament contrast and weak or
moderate values of surface wind speed, where a variable wind speed
exponent is employed to maximize these measures.

Three locations of recent North Atlantic right whale ({\it Eubalaena
  glacialis}) sightings in the Gulf of St.~Lawrence are sampled
between 2008 and 2020 by 324 Radarsat-2 SAR scenes and 10-m wind speed
from the ERA5 reanalysis.  The inverse relationship between SAR
contrast magnitude and wind speed is quantified, and a reduced
correlation is obtained for all three domains when SAR contrast is
weighted by wind speed to the power of 0.8.  A more uniform emphasis
on ocean surface structure within a SAR scene, or across multiple
scenes, can thus be considered in the search for prey aggregations of
the North Atlantic right whale.

\vspace{0.3in}
\noindent\rule{\textwidth}{0.5pt}

\newpage
\begin{multicols}{2}
\normalsize

\section{Introduction}

Ecosystem studies in the Gulf of St.~Lawrence (GSL) are well suited
to address the stock and supply of zooplankton {\it Calanus} spp., the
preferred prey of the endangered North Atlantic right whale (NARW),
{\it Eubalaena glacialis} \citep{Plourde_etal_2019,
  Sorochan_etal_2019, Gavrilchuk_etal_2020, Brennan_etal_2021}.  For
an animal the size of a grain of rice, however, one of the best known
indications of its local aggregation (i.e., only dense aggregations of
{\it Calanus} offer a net energy surplus to a right whale) is the
right whale itself \citep{Baumgartner_etal_2007}.  Observations of
right whales suggest that mothers provide foraging patterns for their
young and sensory perception is employed while feeding.  In addition,
\citet{Kenney_etal_2001} provide a number of hypotheses about how
right whales might discover {\it Calanus} aggregations beyond their
immediate locale.

\end{multicols}

\begin{figure}[hbt]
  \centering
  \includegraphics[width=0.5\textwidth]{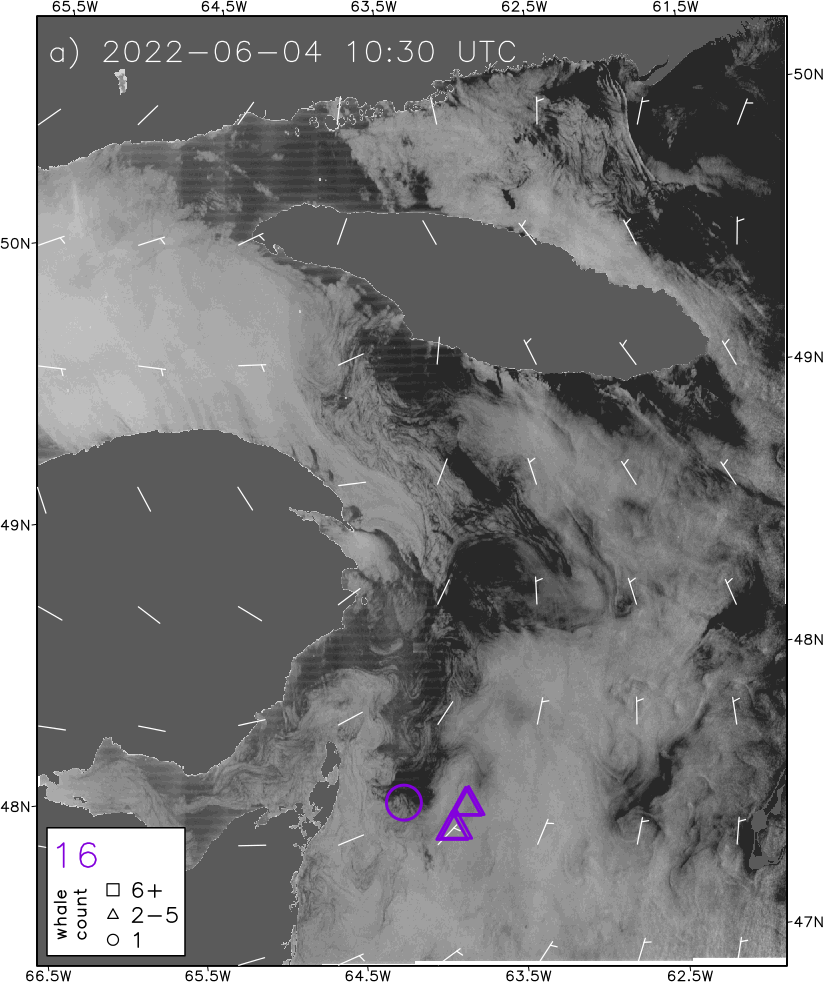}
  \hspace{0.0in}
  \begin{minipage}[b][0.42\textheight][s]{.4\textwidth}
    \centering
    \includegraphics[width=0.9\textwidth]{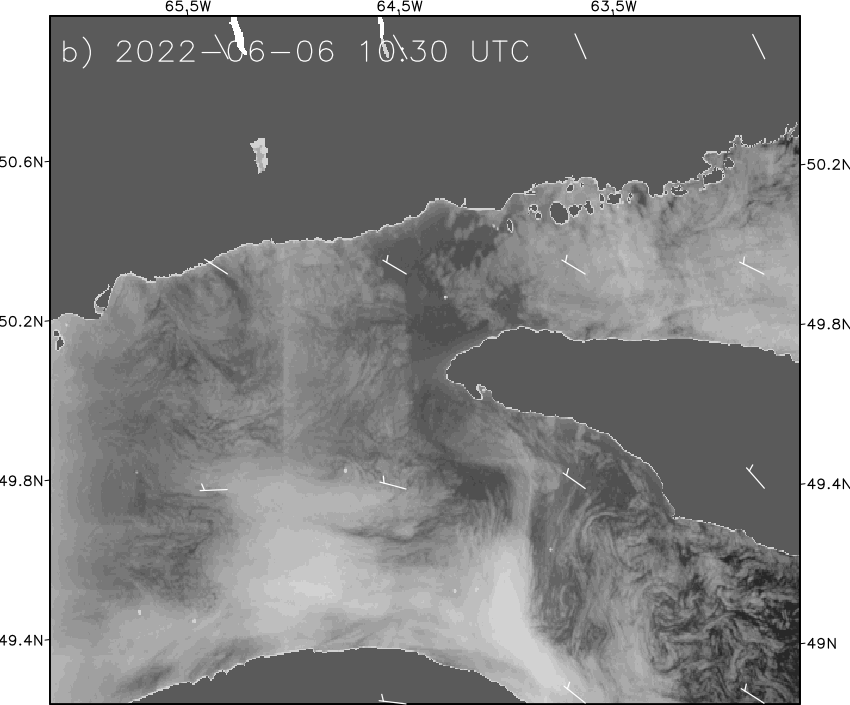}
    \vfill
    \includegraphics[width=0.9\textwidth]{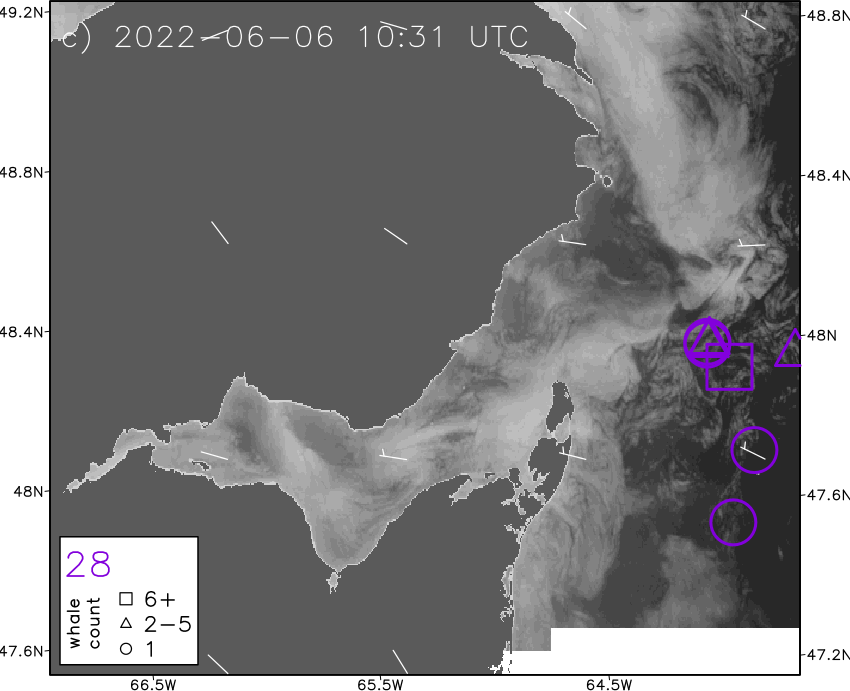}
  \end{minipage}
  \caption{Synthetic aperture radar scenes from descending passes of
    the a)~Radarsat Constellation Mission (RCM-1) and b,c)~Sentinel
    (S-1a) satellites during a quiescent period of weak winds over the
    Gulf of St.~Lawrence.  Shown are normalized radar cross section
    ($\sigma_\circ$) at 10:30~UTC on June~4 (left) and June~6 (right),
    2022.  Dark and light shading denote small and large
    $\sigma_\circ$ (or smooth and rough ocean surfaces), respectively.
    Purple marks are sightings of NARW groups on the corresponding day,
    with the total number of individuals at lower left.  Wind barbs are
    interpolated from the ERA5 reanalysis, with half barbs denoting
    5~ms$^{-1}$.}
  \label{fig00}
\end{figure}

\begin{multicols}{2}

A synthesis of {\it Calanus} spp.\ life cycle and expected biomass
distributions in the GSL is given by \citet{Sorochan_etal_2021}, who
highlight the need to explore NARW prey aggregation at depth, as well
as transient aggregation near the surface, in regions delimited by
tidal mixing and freshwater pulses, and in convergent circulations in
the upper mixed layer (their Fig.~5a).  As with aggregation cues in
complementary high-resolution satellite images
(e.g.~\citealt{Basedow_etal_2019, Barthelmess_etal_2021}), the
interpretation of surface roughness patterns in synthetic aperture
radar (SAR) images is well developed \citep{Holt_2004}.  An historical
account is given by \citet{Munk_etal_2000}, who address the dynamics
of O[10-km] cyclonic eddies, in which coherent filaments at the ocean
surface can be seen to evolve.  Such developments motivate the
question of whether ocean currents, fronts, waves, and eddies that are
revealed in SAR roughness might be helpful to understand NARW prey
aggregation.

Examples of roughness patterns that bracket two days of weak winds
over the GSL are shown in Fig.~\ref{fig00}.  Dark and light shading
denotes smooth and rough ocean surfaces, respectively.  Filaments of
all shades prevail at the smallest scales, with a transverse
separation between filaments of O[10~km] or less.
\citet{McWilliams_etal_2009} demonstrate that a symmetric boundary
layer evolution, organized by an initially frontogenetic (e.g., eddy)
circulation, can result in rapid downwelling.  This is marked in SAR
scenes by darker lines that are thin relative to adjacent lighter
lines (e.g., east of 64.5$^\circ$W).  Such contrasts can also be
enhanced by an accumulation of surfactants (active surface materials)
of terrestrial or atmospheric origin, with sources in the water column
that include bacteria, plankton, fish, and crude oil seepage
\citep{Espedal_etal_1998, Hamilton_etal_2015, Kurata_etal_2016}.

Large scale patterns of surface stress are also apparent in
Fig.~\ref{fig00}.  Dark regions lack the O[1-10-cm] roughness elements
that cause radar backscatter, and indicate wind speeds of less than
about 2~ms$^{-1}$ \citep{Holt_2004}.  However, the southern boundary
of this smooth region also seems to bound a precursor watermass
extension into the southern GSL, close to where 16~NARWs were sighted
(Fig.~\ref{fig00}a).  Further upstream, a bright region along the
north shore of the Gasp\'{e} Peninsula (cf.~Fig.~\ref{fig01}) suggests
a relatively warm and fresh surface layer with enhanced atmospheric
coupling (e.g., northward lines in Fig.~\ref{fig00}a are likely
offshore wind streaks).  Within a few days (Fig.~\ref{fig00}b), this
large scale roughness contrast extends toward groups of NARWs further
north.  Along the boundaries of the precursor and subsequent
watermasses, surface current convergence seems to exist, although as
might be expected \citep{Espedal_etal_1998, Munk_etal_2000}, contrasts
that bound the brighter region are not as robust as those that bound
the darker region of Fig.~\ref{fig00}.  The possible role of
surfactants is discussed further in Section~5.

\subsection{Radar modelling}

This quiescent period of weak winds is exemplary of an upper ocean
current response captured by SAR snapshots.  \citet{Rascle_etal_2017,
  Rascle_etal_2020} further emphasize that ocean currents modulate
waves and wave breaking, and in turn, surface roughness over a wide
range of environmental conditions, including at wind speeds of up to
about 10~ms$^{-1}$.  A theoretical basis for this is given by radar
models that are well suited to address dependencies on viewing and
environmental conditions.  \citet{Kudryavtsev_etal_2012a,
  Kudryavtsev_etal_2012b} provide a unified framework of roughness
{\it contrast}, which for measurements of SAR normalized radar cross
section ($\sigma_\circ$), is defined as
\begin{equation}
  \frac{\sigma_\circ - \overline{\sigma_\circ}}{\overline{\sigma_\circ}}.
  \label{contrasts}
\end{equation}
The overbar denotes a smoothing to lower resolution.  Not only is
roughness contrast an effective method of highlighting filaments of
all shades (cf.~\citealt{Young_etal_2008}), but the corresponding
radar imaging model (RIM; \citealt{Kudryavtsev_etal_2005}) provides a
functional expression that varies linearly with ocean current
convergence, and in weak winds, depends nonlinearly on surfactants
(e.g., in terms of thickness or elasticity).  Contrast also depends
nonlinearly on wind speed, in that it varies almost linearly with the
inverse of friction velocity squared \citep{Kudryavtsev_etal_2012b}.

The functional form of the RIM model suggests that wind stress and
surface current convergence can be considered somewhat independently
of their prior coupled evolution.  In other words, SAR snapshots like
Fig.~\ref{fig00} might capture a similar transient pattern of ocean
current contrasts in stronger or weaker winds, except that the
contrasts themselves are weaker or stronger, respectively.  It is then
possible to consider an adjustment that offers more uniformity in
filament contrast, both within an individual SAR scene and for as many
scenes as are available, say, over the GSL.  This semblance of
uniformity may be helpful to refine the search for transient {\it
  Calanus} aggregations \citep{Kenney_etal_2001, Sorochan_etal_2021}.
Adjusted SAR contrasts can then be interpreted preferentially in terms
of ocean current convergence, or perhaps more accurately, as current
deformation by a combination of convergence and strain
\citep{Rascle_etal_2014}.

\subsection{Measurement modelling}

We introduce measurement modelling as complementary to
process-oriented approaches like radar or ecosystem modelling.  It is
well suited to assess the performance of a SAR contrast adjustment on
wind speed, even though a complex dependence on sea state (and other
variables) is expected.  On one hand, the construction of a
measurement model does not specify the processes of interest, although
our assessment can be considered {\it specific} to the choice of GSL
measurements.  On the other hand, a process model (e.g.,
\citealt{Kudryavtsev_etal_2012b}) targets a {\it general} relationship
that is often consistent with a broad set of measurements, but again
by construction, complex processes must be specified.  Formally, ours
is also called an observational study \citep{Cochran_1972} that
employs correlation measures to identify \citep{Palacios_etal_2013,
  Edelmann_etal_2021} and adjust for measurement dependence.

Using a simple measurement model, we propose to examine a partial, and
therefore nonlinear, relationship that focuses on right whale foraging
habitat and on filaments that are identified as contiguous SAR
contrasts (cf.~\citealt{Young_etal_2008}).  Following
\citet{Szekely_Rizzo_2009} and \citet{Danielson_etal_2020a}, an
attempt is made to quantify the linear and nonlinear components of the
SAR contrast and wind speed relationship.  Section~3 describes an
identification of filaments and our dependence measures.  The
relationship between contiguous contrasts and wind speed is examined
in Section~4, where an adjustment is applied.  Biophysical and
statistical aspects are discussed in Section~5 and conclusions are
given in Section~6.

\end{multicols}

\begin{figure}[hbt]
  \centering
  \includegraphics[width=0.8\textwidth]{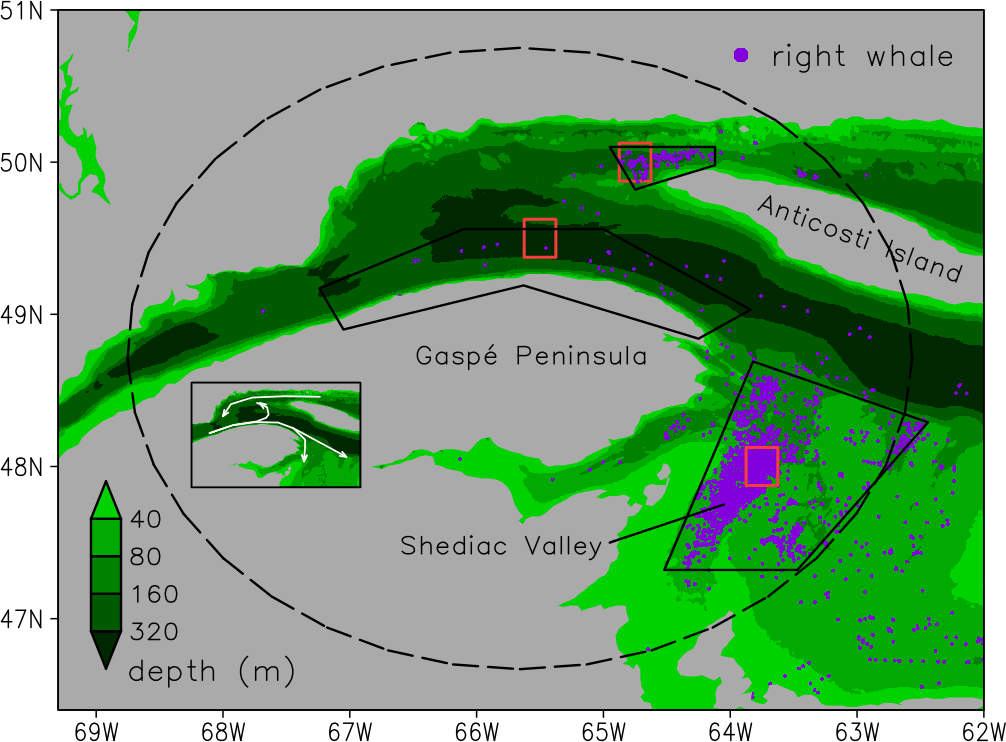}
  \caption{Location of 4900~visual sightings of the North Atlantic
    right whale ({\it Eubalaena glacialis}; purple dots) between 2015
    and 2020, along with bathymetry (green shading) of the western
    Gulf of St.~Lawrence.  Solid black lines encompass the Anticosti
    (northern), Gasp\'{e} (western), and Shediac (southern) domains
    (see text).  Red lines encompass the corresponding ERA5 10-m wind
    speed gridboxes.  White arrows in the inset show the circulation
    pattern of the western Gulf.  Note that sightings are a function
    of search effort, with uneven coverage in space and time.}
  \label{fig01}
\end{figure}

\begin{multicols}{2}

\section{Data}

This section describes right whale sightings for 2015-2020, and SAR
and wind data for 2008-2020, for three subdomains in the Gulf of
St. Lawrence (GSL).  Significant shifts in prey availability during
the 2010s \citep{Sorochan_etal_2019} have led the extant NARW
population away from their traditional foraging habitats in the Gulf
of Maine and Scotian Shelf and toward shallower regions of the GSL,
where they were sighted only occasionally prior to 2013
\citep{Kenney_2019}.  Since 2015, NARWs have been sighted between May
and November, more frequently along the Shediac Valley in the
southwestern GSL, and in a small region just northwest of Anticosti
Island (Fig.~\ref{fig01}).  As part of a broad cyclonic circulation in
the Gulf \citep{Koutitonsky_Bugden_1991, Lavoie_etal_2016}, the
Anticosti sightings are furthest upstream (northern domain in
Fig.~\ref{fig01}) and flow is generally westward along the north shore
until it merges with the Gasp\'{e} Current (elongated western domain).
The relatively fresh St.\ Lawrence estuary outflow also feeds this
coastal current, as it flows around the Gasp\'{e} Peninsula.  The
Gasp\'{e} Current is known to transport {\it Calanus} spp., with
periodic intrusions onto the Shediac Valley (the southern domain;
\citealt{Brennan_etal_2021}).  The 2019 warm season is notable for
having high St. Lawrence river runoff and good SAR coverage.

\subsection{Whale Sightings}

Typical locations of GSL prey aggregation are taken from two sources
of whale sightings.  The North Atlantic Right Whale Consortium (NARWC)
sightings database \citep{Kenney_2019, NARWC_2021} includes
3265~aerial survey and vessel-of-opportunity right whale sightings
(either of a whale group or an individual whale) from 2015 onward in
the GSL.  A complementary Fisheries and Oceans Canada (DFO) database
contains 2586~sightings of right whales \citep{DFO_TeamWhale_2020},
although this excludes a subset of aerial surveys from the most recent
years.  Some DFO sightings might have been of dead whales, but this is
only recorded in the NARWC data (which reduces the NARWC data from
3265 to 3188).  The DFO sightings are reported in local time and UTC
(NARWC sightings are in UTC).

The NARWC and DFO datasets each contain identical sightings separated
by only a few minutes and less than a kilometer\footnote{Information
  about right whale presence and absence is generally sparse, but
  opportunistic sightings are more frequent closer to shore and during
  the summer months.  Sighting {\it effort} also depends on observing
  conditions (e.g., visibility depends on sea state, fog,
  precipitation, and time of day).  Effort increased after 2016 but
  was affected by COVID-19; no effort corrections are employed here.}.
This highlights that the same observer, or multiple observers, may
report on the same whale more than once (and that some whales are
counted more often than others, either on the same day or on different
days).  This observing-effort duplication is retained.  Unrelated to
effort, the NARWC and DFO sightings are also partially overlapping, so
duplicates are omitted if two sightings of the same number of whales
occurs within 24~h and 0.01$^\circ$ latitude and longitude.  This
reduces the combined number of sightings by 874.  The remaining
4900~sightings of a group or individual right whale (1712 and 3188;
Fig.~\ref{fig01}) corresponds to 7591~sightings of individuals.  The
counts per year are listed in Table~\ref{tab01}, with the caveat that
2019 and 2020 sightings are preliminary.

\end{multicols}

\addtolength{\tabcolsep}{-2pt}
\begin{table}[ht]
  \begin{center}
    \begin{tabular}{|c|c|c|cccc|ccc|}
      \hline
      Year & NARW Sightings & SAR Scenes & \multicolumn{4}{c|}{Scenes with NARWs} & \multicolumn{3}{c|}{Sightings in Scenes} \\
      \hline
      2015 &    114/343    &     86     & & &    2 & (2\%)       & &    2/3   &  (2\%/1\%) \\
      2016 &    100/159    &     53     & & &    3 & (6\%)       & &    6/9   &  (6\%/6\%) \\
      2017 &   1151/2118   &     83     & & &   11 & (13\%)      & &  165/278 & (14\%/13\%) \\
      2018 &   1622/2913   &    105     & & &    4 & (4\%)       & &   31/33  &  (2\%/1\%) \\
      2019 &   1904/2049   &    188     & & &   31 & (16\%)      & &  669/705 & (35\%/34\%) \\
      2020 &      9/9      &     68     & & &    0 & (0\%)       & &    0/0   &  (0\%/0\%) \\
      \hline
    \end{tabular}
  \end{center}
  \vspace{-0.2in}
  \caption{Number of Gulf of St. Lawrence NARW sightings (by
    group/individual), and SAR scenes per year between 2015 and 2020
    (the 2019 and 2020 sightings are preliminary).  Included are the
    number of scenes with right whales, and the number of right whale
    sightings (by group/individual) collocated with those scenes, with
    fractions of their total in brackets.  A sighting is considered to
    be temporally collocated with a SAR scene if both occur on the
    same day.  Coverage by SAR (of dashed circle in Fig.~\ref{fig01})
    is from May to December and includes 11~(2008), 24~(2009),
    31~(2010), 99~(2011), 63~(2012), 68~(2013), and 62~(2014) scenes
    prior to 2015.}
  \label{tab01}
\end{table}

\begin{multicols}{2}

\subsection{Synthetic Aperture Radar (SAR)}

The Radarsat-2 C-band (5.3-cm) SAR has been operating since May 2008
in a polar Earth orbit about 800~km above the surface.  It is capable
of transmitting and receiving in horizontal and vertical (H and V)
polarization at incidence angles between 20$^\circ$ (near range) and
49$^\circ$ (far range).  GSL coverage is available at a pixel
resolution of 100~m or smaller over swaths of more than 300~km.  We
consider all acquisitions during the ice-free months of May through
December of 2008-2020 that overlap with the dashed circle in
Fig.~\ref{fig01}.  This yields 941~SAR scenes using transmit and
receive with the same polarization (i.e., HH or VV), and all but
three employ a narrow or wide ScanSAR beam mode.  Although SAR
coverage of the GSL varies somewhat by year (Table~\ref{tab01}),
sightings coverage is notable in 2017 and 2019, with 2019 having the
most sightings (1904), the best GSL coverage (188 scenes), and the
best sightings coverage (35\%).

\begin{figure}[hbt]
  \centering
  \includegraphics[width=0.8\textwidth]{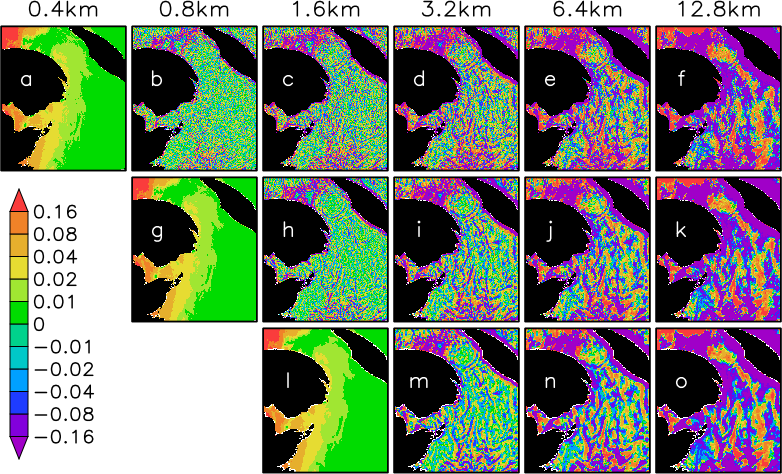}
  \caption{Normalized radar cross section ($\sigma_\circ$) and SAR
    contrast equation (\ref{contrasts}) in linear units and at different pixel
    resolutions for a Radarsat-2 scene (HH) acquired over the Gulf of
    St.\ Lawrence on July 26, 2019 (21:59~UTC).  Shown are
    $\sigma_\circ$ at a)~0.4-km resolution and contrast at scales up
    to b)~0.8~km, c)~1.6-km, d)~3.2-km, e)~6.4-km, and f)~12.8-km
    resolution.  Similarly, $\sigma_\circ$ is shown at g)~0.8~km with
    contrast to h-k)~1.6-km through~12.8-km resolution, and at
    l)~1.6-km with contrast to m-o)~3.2-km through~12.8-km resolution,
    respectively.  Values are given by the colourbar on the left.}
  \label{fig02}
\end{figure}

We focus on a subset of scenes that provide coverage of the Anticosti,
Gasp\'{e}, and Shediac domains (Fig.~\ref{fig01}) between mid-May and
mid-August (i.e., when whales are foraging, but winds are weak and
filaments are easier to identify).  This subset has 177, 241, and 237
scenes, respectively (a total of 324 scenes).  Initial SAR processing
involves reducing the resolution of normalized radar cross section
($\sigma_\circ$) by a smoothing operator that halves resolution on
each pass \citep{Koch_2004}.  This yields scenes at 100-m, 200-m,
400-m, 800-m, 1600-m, 3200-m, 6400-m, and 12800-m resolution (e.g.,
Fig.~\ref{fig02}a,g,l).  At 800-m resolution, we also perform an
preliminary masking of each scene using the CMOD5 VV ocean wind model
\citep{Hersbach_etal_2007}, with the incidence-angle polarization
ratio of \citet{Zhang_etal_2011} for HH scenes.  This is done to
partially remove features unassociated with filaments, but to retain a
border around any patterns of interest within scenes like
Fig.~\ref{fig00}.  As illustrated in the next section
(Figs.~\ref{fig02} and~\ref{fig03}), this preliminary masking assumes
bounds on $\sigma_\circ$ corresponding to a 1-ms$^{-1}$ wind directed
parallel to the satellite track, and a 15-ms$^{-1}$ wind directed
toward the satellite.

\end{multicols}

\begin{multicols}{2}

\subsection{Surface Wind Analysis}

A SAR contrast dependence on wind speed is somewhat simpler to
interpret if wind speed does not depend on (or assimilate) SAR
backscatter.  Thus, zonal and meridional wind analyses at 10~m above
the surface are taken from the fifth European Centre for Medium-Range
Weather Forecasting Reanalysis (ERA5; \citealt{Hersbach_etal_2020}).
This reanalysis employs a spectral atmospheric forecast model and a
sequential data assimilation system with 137~vertical levels and an
effective horizontal resolution of 31~km.  The four-dimensional
variational (4D-Var) analysis is performed over successive 12-h
periods.  Bias adjustment of selected observations accommodates
systematic differences between the model and observations
\citep{Dee_2005}.  The global atmosphere, land surface, and ocean
surface wave evolution is given at hourly intervals from 1950 onward.

The 10-m wind analyses are obtained from an archive that is resampled
at 0.25$^\circ$, and three sampled locations (red boxes in
Fig.~\ref{fig01}) are taken as representative ERA5 surface wind
estimates for the Anticosti, Gasp\'{e}, and Shediac domains.  (It is
convenient to employ a single gridbox for each domain, as this
facilitates measurement model solutions that employ samples at
adjacent hours.)  We focus on relatively strong SAR contrasts that are
indicative of ocean current convergence, perhaps enhanced by the
presence of surfactants \citep{Munk_etal_2000, Rascle_etal_2020}.
Wind speed is thus restricted to 1-10~ms$^{-1}$ when collocating with
SAR contrast in each domain.  Values of 1-3~ms$^{-1}$ are retained
here to include strong contrasts (i.e., ERA5 weak wind estimates are
approximate; cf.~Fig.~\ref{fig00}).

\section{Methods}
\label{methods}

Aggregations of NARW prey may be only partly localized within a given
SAR scene, but the apparent dynamical connections in surface roughness
provide more focus for our search.  Observations and models motivate a
SAR wind speed adjustment, but here, we seek an adjustment that is
specific to our choice of two measurements (i.e., coherent contrasts
and ERA5 wind speed in the Fig.~\ref{fig01} domains).  First, we
confirm that contrast magnitude depends on wind speed.  Next, because
this partial relationship is expected to be nonlinear (e.g.,
\citealt{Kudryavtsev_etal_2012b}), we gauge a dependence on wind speed
to some power.  The power that yields the strongest dependence is
taken as appropriate.  Our experimental design and proposed adjustment
rely on a)~an identification of coherent SAR contrasts, and
b)~measures of the strength of their wind speed dependence.

\end{multicols}

\begin{figure}[hbt]
  \centering
  \includegraphics[width=0.9\textwidth]{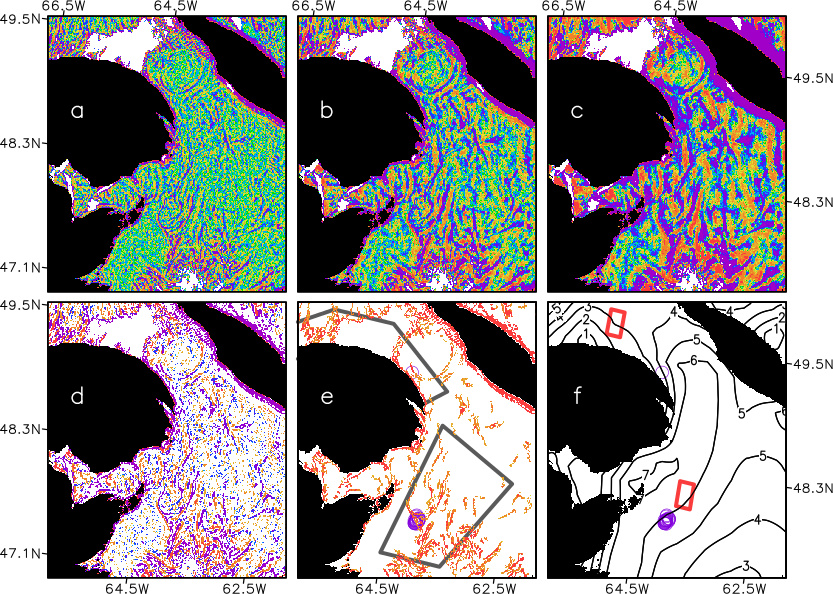}
  \caption{Filament processing and ERA5 wind speed, where (a,b,c)~are
    the same as Fig.~\ref{fig02}h,i,j, but include SAR wind speed
    masking (see text).  Also shown are d)~an average of (a,b,c) for
    contrast values of the same sign that are greater than 0.3 in
    magnitude, e)~contrast magnitude for filaments defined by the
    contiguous values of (d) that cover a distance of at least 10~km,
    and f)~ERA5 10-m wind speed at 1-ms$^{-1}$ intervals.  Values are
    in linear units in (a-e) with colours as in Fig.~\ref{fig02}.
    Included in (e) and (f) are right whale sightings on this day,
    July~26, 2019 (purple circles), and e)~SAR (grey) and f)~ERA5
    (red) domains of Fig.~\ref{fig01}.}
  \label{fig03}
\end{figure}

\begin{multicols}{2}

\subsection{Filament Coherence}
\label{filament}

In assessing ways to identify roughness features in SAR imagery,
\citet{Young_etal_2008} note that ``manual phenomena
identification...is easier following feature extraction...and easier
still following pixel aggregation.''  Here, SAR contrast in equation
(\ref{contrasts}) is taken to capture (or extract) the features of
interest, but in support of a search (manual or otherwise) for {\it
  Calanus} hotspots, coherence (or pixel aggregation) can also be
addressed.  Although 286 (88\%) of the 324~GSL SAR scenes employ VV
polarization, a separate HH wind speed adjustment is not considered
here because local SAR measurement contrast in equation
(\ref{contrasts}) is given as an HH or VV ratio.  We take coherence to
be contiguous contrasts that would nearly span a submesoscale eddy
\citep{Munk_etal_2000}, by a distance of at least 10~km.  Following
\citet{Kudryavtsev_etal_2012b}, $\sigma_\circ$ and
$\overline{\sigma_\circ}$ can be averages over moving windows of, say,
250-m and 25-km respectively.  Although such boxcar averages can
include a wide range of finescale structure (i.e., roughly equivalent
to the scales between our 100-m and 6.4-km nominal pixel resolutions;
\citealt{Koch_2004}), an initial examination of a range of resolution
brackets helps to determine the shape and extent of the filaments of
interest.

There are 28~possible combinations of $\sigma_\circ$ and
$\overline{\sigma_\circ}$ for resolutions between 100~m and 12800~m.
A subset of 12~combinations is shown in Fig.~\ref{fig02} for a SAR
scene acquired on July 26, 2019.  As expected, large scale contrast
patterns vary mainly with $\overline{\sigma_\circ}$, so column panels
are similar.  However, $\overline{\sigma_\circ}$ at 800-m
(Fig.~\ref{fig02}b) and 12800-m (Fig.~\ref{fig02}f,k,o) resolution
seem to capture not enough or too much of this large scale pattern,
respectively.  Instead, we opt to define filaments where there is
agreement in contrast among, for example, Fig.~\ref{fig03}a,b,c.  That
is, we consider $\sigma_\circ$ at 800-m, and $\overline{\sigma_\circ}$
at 1600-m, 3200-m, or 6400-m pixel resolution.  By agreement, we mean
an average of these three contrast values, but only where all are
positive or negative and larger in magnitude than 0.3.  This agreement
among three overlapping resolution brackets serves to isolate
contiguous contrasts (Fig.~\ref{fig03}d), from which filaments are
identified that span a distance of at least 10~km (Fig.~\ref{fig03}e).
To compare with representative values of ERA5 wind speed
(Fig.~\ref{fig03}f red boxes), we obtain a single value of contrast
magnitude for each of the three GSL domains by averaging on the
overlap with each SAR scene.  Because overlap values are set to zero
outside filaments, it is only the values inside filaments that make a
nonzero contribution to domain averages.

Figure~\ref{fig03} helps to motivate an adjustment for wind speed
across a single SAR scene.  Twenty-one right whale sightings are
clustered in Fig.~\ref{fig03}e (purple circles) along filaments that
appear to bound a southward intrusion of the Gasp{\'e} Current into
the Shediac Valley.  Further upstream, between the Gasp{\'e} Peninsula
and Anticosti Island is the signature of a large anticyclonic
excursion of the Gasp{\'e} Current (i.e., an Altika pass on this day
indicates a positive sea surface height anomaly; not shown).  The
anticyclonic excursion and Shediac intrusion are collocated with
relatively strong wind speed (Fig.~\ref{fig03}f) and are weakly
delimited by filaments.  On the other hand, further south of the
whales is a cyclonic eddy surrounded by extensive filaments that are
partially masked (Fig.~\ref{fig03}a,b,c).  Similarly, part of the
Gasp{\'e} domain is masked where wind speed is weak.  Although
filaments are extensive where $\sigma_\circ$ is reduced, the search
for {\it Calanus} aggregations is likely to be more successful in the
presence of whales \citep{Kenney_etal_2001, Baumgartner_etal_2007}.
In this case, an adjustment that emphasizes the contrasts in strong
wind, relative to contrasts in weak wind, seems reasonable.

\subsection{Dependence Measures}

\citet{Szekely_etal_2007} and \citet{Szekely_Rizzo_2009} formulate a
measure of dependence called distance correlation that is sensitive to
nonlinear and nonmonotone measurement covariance, and thus extends and
complements Pearson correlation.  Whereas Pearson correlation is known
mainly as a measure of linear dependence, it is also possible to
formulate a wavelike measurement model that accommodates both linear
and nonlinear dependence \citep{Danielson_etal_2018,
  Danielson_etal_2020a}.  As noted in Section~5, autocorrelation
provides a basis for wavelike model solutions.  For SAR contrast
magnitude ($C$) and ERA5 wind speed ($U$) in the domains of
Fig.~\ref{fig01}, the model, variance, and cross-covariance components
are
\begin{eqnarray}
  \begin{array}{r} C\color{white}{)} \color{black} \\ U\color{white}{)} \color{black} \\ Var(C) \\ Var(U) \\ Cov(C,U) \end{array}
  \hspace{-0.05cm}
  \begin{array}{c}      =     \\      =     \\        =       \\     =     \\     =     \end{array}
  \hspace{-0.05cm}
  \begin{array}{c}            \\   \alpha_U \\                \\           \\     \color{white}{)}      \end{array}
  \hspace{\negicm}
  \begin{array}{c}            \\      +     \\                \\           \\     \color{white}{)}      \end{array}
  \hspace{-0.50cm}
  \begin{array}{r}      t     \\  \beta_U t \\    \sigma_t^2   \\ \beta_U^2 \sigma_t^2 \\ \beta_U \sigma_t^2 \end{array}
  \hspace{\negicm}
  \begin{array}{c}      +     \\      +     \\         +        \\     +     \\     +     \end{array}
  \hspace{\negicm}
  \begin{array}{c}  \epsilon  \\  \epsilon  \\ \sigma_\epsilon^2 \\ \sigma_\epsilon^2 \\ \sigma_\epsilon^2. \end{array}
  \hspace{\negicm}
  \begin{array}{c}      +     \\      +     \\         +        \\     +     \\     \color{white}{)}      \end{array}
  \hspace{\negicm}
  \begin{array}{c} \epsilon_C \\ \epsilon_U \\    \sigma_C^2    \\ \sigma_U^2 \\     \color{white}{)}      \end{array}
  \label{var}
\end{eqnarray}
This model introduces linear association ($t$), nonlinear association
($\epsilon$), and a lack of association ($\epsilon_C$ and
$\epsilon_U$) as signal-and-noise terms (but whose interpretation is
based on signal), with variance $\sigma_t^2$, $\sigma_\epsilon^2$, and
$\sigma_C^2$ and $\sigma_U^2$, respectively.  Such a framework can be
said to relate measurements with each other, but only by way of what
they both measure.  Thus, $t + \epsilon$ (in $C$) and $\alpha_U +
\beta_U t + \epsilon$ (in $U$) capture the total association between
$C$ and $U$, where $\alpha_U$ and $\beta_U$ are, respectively, an
additive and multiplicative calibration of $t$ in $U$.

Dependence of SAR contrast magnitude ($C$) on ERA5 wind speed ($U$)
can be measured using both Pearson and distance correlation.  A novel
decomposition of Pearson correlation [$\rho = Cov(C,U) / \sqrt{Var(C)
    Var(U)}$] is also permitted by the linear ($\beta_U \sigma_t^2$)
and nonlinear ($\sigma_\epsilon^2$) components of $Cov(C,U)$.
Multiple solutions of each component are given below, based on samples
of ERA5 wind speed at intervals from 1~h to 5~h (cf.~Section~5).  The
two key parameters required to solve equation (\ref{var}) are $\beta_U$ and
$\sigma_t^2$, and for smooth solutions, it is convenient to fix one of
these.  Although \citet{Danielson_etal_2018} fix $\beta_U$, here we
fix $\sigma_t^2$ to the reverse linear regression value of $Cov^2(C,U)
/ Var(U)$.

\end{multicols}

\begin{figure}[hbt]
  \centering
  \includegraphics[width=0.9\textwidth]{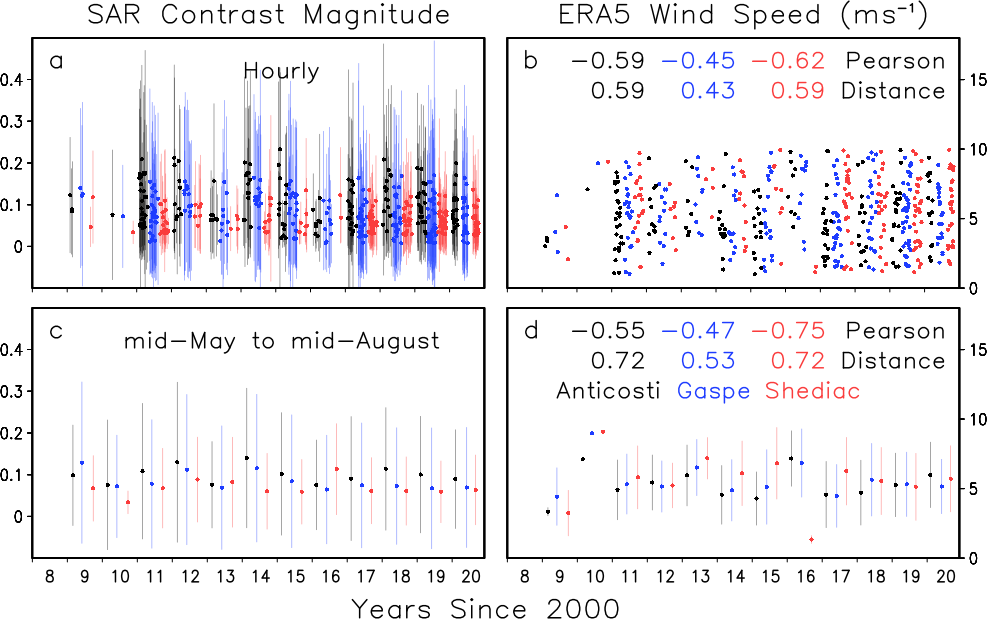}
  \caption{Mid-May to mid-August values of a,c)~SAR contrast magnitude
    ($C$ in linear units) and b,d)~ERA5 wind speed ($U$ in ms$^{-1}$)
    during 2008-2020 for the Anticosti (black), Gasp{\'e} (blue), and
    Shediac (red) domains of Fig.~\ref{fig01}.  The Anticosti and
    Shediac values are shifted by 3.5~months earlier and later,
    respectively.  Shown are averages a,b)~for each domain overpass
    (hourly bin interval) and c,d)~for all overpasses between mid-May
    and mid-August (annual interval; dots with lines above and below
    of one standard deviation), and the corresponding Pearson and
    distance correlation values for each domain and time interval.}
  \label{fig04}
\end{figure}

\begin{multicols}{2}

Dependence within each domain of Fig.~\ref{fig01} is examined by
correlating $C$ and $U^x$, where $x$ is a variable exponent in the
range [-5,5].  The value $x^*$ that yields the strongest correlation
is then taken as an appropriate wind speed adjustment.  Because the
relationship is inverted (e.g., strong contrast tends to occur in weak
winds), we multiply each of three SAR contrast values by $(V /
6)^{x^*}$, where $V$ is wind speed interpolated across the scene
(i.e., applied to contrast at 800-m resolution, with
$\overline{\sigma_\circ}$ at 1600-m, 3200-m, and 6400-m pixel
resolution, including outside the domains of Fig.~\ref{fig01}).  Also,
because neither Pearson nor distance correlation depend on a
multiplicative calibration of $C$ \citep{Szekely_etal_2007}, we
include the arbitrary normalization $(1 / 6)^{x^*}$ as part of this
adjustment, so contrast values at wind speeds of 6~ms$^{-1}$ are
unchanged.

\section{Results}

Right whale foraging in the GSL can extend into the fall
\citep{NARWC_2021, DFO_TeamWhale_2020}, but we focus on SAR scenes
during mid-May to mid-August, when winds are lighter and {\it Calanus}
near the surface may be more dense \citep{Brennan_etal_2021,
  Sorochan_etal_2021}.  It is easier to pick out a dependence on ERA5
wind speed in the seasonal averages (Fig.~\ref{fig04}c,d) than for the
individual overpasses (Fig.~\ref{fig04}a,b).  For example, during
years of good SAR coverage, wind speed in the Shediac domain tends to
be larger (red dots in Fig.~\ref{fig04}d) and contrast magnitude
smaller (Fig.~\ref{fig04}c) than in other domains.  As expected,
Pearson correlation is uniformly negative.  Distance correlation,
which only takes on values between zero and one (i.e., where zero
indicates independence), is slightly smaller in the Gasp{\'e} domain,
perhaps in part because its ERA5 gridbox is less representative of
wind speed throughout the corresponding SAR domain (Fig.~\ref{fig01}).
Nevertheless, Pearson and distance correlation both seem consistent
with observations \citep{Espedal_etal_1998, Munk_etal_2000} and model
predictions \citep{Kudryavtsev_etal_2012a, Kudryavtsev_etal_2012b} of
a filament contrast dependence on wind speed.

\end{multicols}

\begin{figure}[hbtp]
  \centering
  \includegraphics[width=0.9\textwidth]{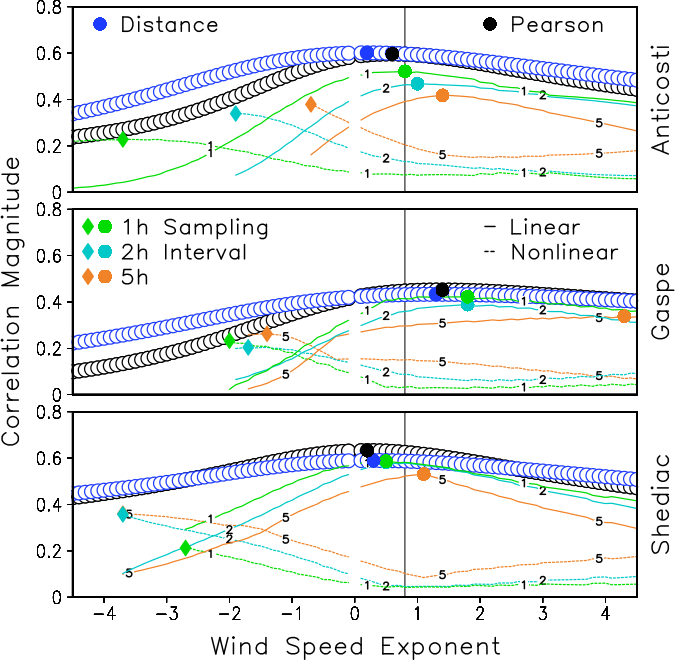}
  \caption{Dependence of Radarsat-2 SAR contrast magnitude ($C$) on
    ERA5 wind speed ($U^x$) as a function of wind speed exponent ($x$,
    abscissa) for the Anticosti (top), Gasp{\'e} (middle), and Shediac
    (bottom) domains.  Shown are the magnitude of distance correlation
    (blue circles) and Pearson correlation (black circles), with
    linear (solid lines) and nonlinear (dashed lines) components of
    Pearson correlation based on solutions of equation (\ref{var})
    that employ four additional samples of ERA5 wind speed at 1-h
    (green), 2-h (cyan), and 5-h (orange) intervals.  Filled circles
    and diamonds are maximum values.  The vertical line at $x = 0.8$
    marks a proposed adjustment.}
  \label{fig05}
\end{figure}

\begin{multicols}{2}

Figure~\ref{fig05} depicts ($C$, $U^x$) correlation magnitude as a
function of wind speed exponent ($x$) for the hourly binned SAR and
ERA5 data of Fig.~\ref{fig04}a,b.  Excluded are values at zero
exponent and any component solutions that are unavailable.  Across all
exponents, distance correlation seems more similar in value (i.e.,
flatter) than the Pearson correlation.  Most correlation minima occur
at large negative exponents.  Maxima in Pearson and distance
correlation (filled circles) are similar to the tabulated values of
Fig.~\ref{fig04}b (at an exponent of one), but typically occur at an
exponent of less than one.  At least for the Gasp{\'e} and Shediac
domains, maxima in Pearson correlation exceed that of distance
correlation to a degree expected of bivariate normal data \citep[\
their Fig.~1]{Szekely_etal_2007}.  This is consistent with small
skewness of the $U^x$ distributions for $x$ in the range [0,2] (not
shown).

Away from their maxima, distance correlation is larger in magnitude
than Pearson correlation, which seems consistent with a greater
sensitivity to nonlinear and nonmonotonic dependence
\citep{Szekely_etal_2007}.  Solutions of equation (\ref{var}) at fixed
$\sigma_t^2$ are not always available, but where there is a nonlinear
contribution to Pearson correlation (filled diamonds), the larger
value of distance correlation also suggests a nonlinear dependence.
Pearson and distance correlation thus seem quite consistent in this
study.  Both suggest a wind speed adjustment by a small positive
exponent that is close to the dominant linear dependence, but is
shifted slightly toward a secondary nonlinear dependence that is
negative (and moreso for shorter sampling intervals).  Also notable is
that the linear and nonlinear correlation maxima of Fig.~\ref{fig05}
are not directly comparable to the SAR contrast formulae of
\citet{Kudryavtsev_etal_2012b}, but the use of friction velocity
instead of wind speed might facilitate a more direct comparison.

\end{multicols}

\begin{figure}[hbt]
  \centering
  \includegraphics[width=0.9\textwidth]{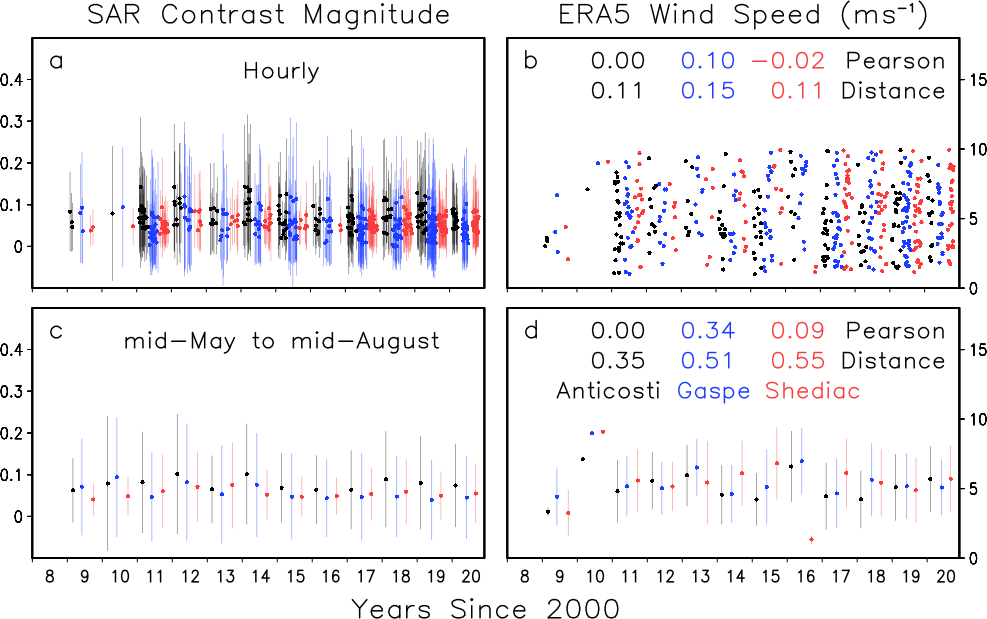}
  \caption{As in Fig.~\ref{fig04}, but following an adjustment of SAR
    contrast magnitude.}
  \label{fig06}
\end{figure}

\vspace{0.1in}
\begin{figure}[hbt]
  \centering
  \includegraphics[width=0.9\textwidth]{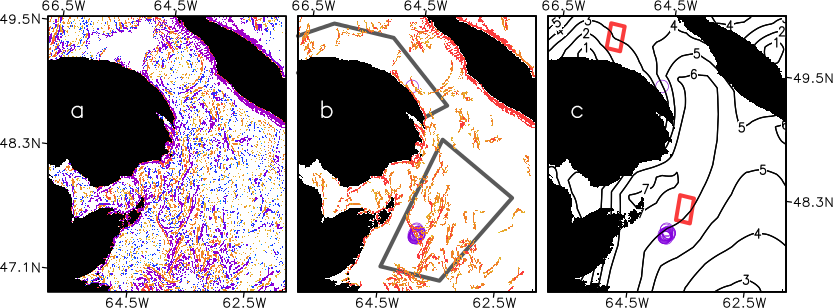}
  \caption{As in Fig.~\ref{fig03}d,e,f, but following an adjustment of
    SAR contrast magnitude.}
  \label{fig07}
\end{figure}

\begin{multicols}{2}

The vertical line at $x = 0.8$ in Fig.~\ref{fig05} provides a
compromise for the three GSL domains.  We multiply SAR contrast by $(V
/ 6)^{0.8}$ to obtain an enhancement or damping, according to whether
the ERA5 wind speed collocation ($V$) is greater than or less than
6~ms$^{-1}$, respectively.  This is done prior to an identification of
coherent filaments (Section~\ref{filament}) and yields a reduction in
the mean and variance of SAR contrast magnitude (Fig.~\ref{fig06}a,c).
Whereas Pearson and distance correlation are reduced to values of less
than 0.2 (Figure~\ref{fig06}b), exponent values greater than 0.8 do
not yield a similarly reduced correlation (not shown).  This suggests
that our adjustment yields a reduction mainly in linear dependence,
but also to some extent in nonlinear dependence.  For the scene on
July 26, 2019 (Fig.~\ref{fig07}), filaments oriented along the
southward intrusion of the Gasp{\'e} Current into the Shediac domain,
while small, also become more numerous.  A favourable reduction in
weak wind contrast at the scene borders is also apparent.  In this
case, what appear to be the transient watermass boundaries of interest
are largely unchanged, but the orientation of the filaments that mark
these boundaries is easier to identify.  Although SAR contrast
variance is still large (Fig.~\ref{fig06}a,b), a simple adjustment
provides more uniformity across individual scenes.

\section{Discussion}

Boundary layer circulations may provide an initial organization of
filaments, which subsequently gather along watermass strainlines to
form extensive connections \citep{Munk_etal_2000, Peacock_Haller_2013,
  Maps_etal_2015}.  \citet{Dong_etal_2021} propose that zooplankton
aggregations can be enhanced by such transient material boundaries.
However, at relatively large scales (i.e., larger than an O[10-km]
eddy), the typical SAR signature of biological aggregation, and
specifically one that depends on {\it Calanus} motility, has yet to be
proposed (i.e., including the ability of zooplankton to maintain their
depth, or otherwise locate themselves within the water column).
\citet{Sorochan_etal_2021} note that aggregation in the horizontal and
vertical are both important, either near the surface or at depth.  In
turn, aggregations that are nearly collocated with SAR filaments would
be expected, as suggested in Fig.~\ref{fig07}, as are aggregations
delineated by, but not necessarily collocated with, coherent watermass
boundaries, like those suggested in Fig.~\ref{fig00}.  Despite many
processes within the water column that SAR is sensitive to
\citep{Holt_2004, McWilliams_etal_2009, Shen_etal_2020}, contrast in
equation (\ref{contrasts}) also seems to provide an important spatial
reference in a directed search for NARW prey aggregation.

Processes occurring along filaments and material boundaries are also
key to this search, although \citet{Rascle_etal_2020} emphasize that
we are limited in our ability to distinguish actively convergent and
decaying filaments.  This is needed to identify aggregation processes
from space (i.e., to address what aggregation is), but is in part
complementary to the question of how best to employ a given platform
to measure them.  For the scales that we are able to resolve here, a
reduction in the nonlinear dependence on wind speed is proposed.
\citet{Szekely_Rizzo_2009} note that a separation of distance
correlation into linear and nonlinear components is desirable.
Because autocorrelation pertains to the solution of equation
(\ref{var}), which yields this separation in Pearson correlation, a
few added comments seem relevant.  All correlation values in
Figs.~\ref{fig04}-\ref{fig06} are obtained under two temporal
autocorrelation assumptions: beyond 12~h autocorrelation is weak (more
reasonable for SAR contrast than for ERA5 wind speed) and within 12~h
it is strong (more reasonable for ERA5 wind speed than SAR contrast).
The first assumption permits the use of standard correlation
calculations, where it is convenient to treat each timeseries sample
as independent.  We note that a few SAR scenes are acquired within
12~h of each other in each domain of Fig.~\ref{fig04}, but most scenes
are well separated.

The second assumption is the basis for solutions of equation
(\ref{var}), by sampling ERA5 wind speed twice before and after each
SAR overpass.  As in \citet{Danielson_etal_2018}, a wavelike
measurement model can be written involving autocorrelated samples
(i.e., where the autocorrelation is wavelike).  For sampling intervals
of 1~h to 5~h, the outer samples are, respectively, 2~h and 10~h
before and after.  The hourly ERA5 data thus permit solutions of
equation (\ref{var}), and their autocorrelation over 12~h is more
robust, so this assumption also seems reasonable.

Regarding the origin of nonlinearity in SAR contrast, we expect a
confounding (or unobserved) dependence on surfactant properties that
is in addition to a dependence on wind speed.  The presence of
surfactants is also hinted at in the Introduction and a few general
comments seem relevant.  A review of the state-of-the-art for SAR
imaging of surface slicks like crude oil is given by
\citet{Li_etal_2019} and \citet{Huang_etal_2022}.  SAR is an
effective satellite sensor for monitoring of marine surface slicks
because of its high-resolution, large areal coverage, and all-weather,
day or night capability.  Surfactants on the ocean surface tend to
dampen the wind-generated surface short gravity and capillary waves
that are responsible for the radar backscattering.  The dampening is
due to increased viscosity and decreased surface tension, which
significantly reduces the radar backscatter, resulting in dark patches
in SAR images.  A challenge for SAR detection of surface slicks is the
competition from other natural 'look-alike' phenomena, which may also
reduce radar signals thereby resulting in dark areas
(e.g.,~Fig.~\ref{fig00}).  Look-alike phenomena include low-wind
areas, biogenic slicks, rain cells, grease ice or frazil sea ice,
surface current shears, wind-shadow areas near coasts, upwelling
zones, and internal waves \citep{Alpers_etal_2017}.

To identify surface slicks with single-polarization SAR intensity
images, earlier studies used threshold methods and a statistical
classifier \citep{Fiscella_etal_2000}. However, accurate segmentation
of dark spots from the clean ocean surface areas is difficult using a
simple threshold method for image intensity, because of speckle noise
and heterogeneity in the radar backscatter, from near range to far
range.  Using an adaptive threshold methodology in low winds, dark
spot segmentations might contain many look-alikes.  Thus, to improve
the detection of surface slicks, additional prior knowledge is often
needed, such as ocean surface winds etc.\ \citep{Mera_etal_2012}. In
any case, these approaches involve complex processes to achieve
surface slick detection and classification.

Compared to single-polarization SAR data, which gives the intensity of
radar returns, fully polarimetric SAR measures the scattering matrix
of targets, providing amplitude and phase data for each image pixel.
Therefore, quad-polarization SAR can estimate feature parameters based
on polarimetric decomposition theory, which can discriminate different
scattering mechanisms related to surface slicks. For example, fully
polarimetric, C-band SAR has been also used to identify oil spills,
using various polarimetric features like co-polarized phase
difference, etc.\ \citep{Nunziata_etal_2013, Zhang_etal_2011}.
However, quad-polarization SAR is not practical for operational
monitoring because its coverage is too small (25 or 50~km), whereas
single- and dual-polarization SAR is larger (300~km or 500~km).  But
discrimination of surface slicks from look-alikes is sometimes still
not possible, when the wave-damping behaviors are too similar.  Even
if discrimination could be done, however, surfactants would be
expected to contribute to nonlinearity in Fig.\ref{fig05}.

\section{Conclusions}

The presence of NARWs remain among the best indications of high {\it
  Calanus} concentrations, but given a supply of {\it Calanus} in the
Gulf of St. Lawrence (GSL), can we also say that prey aggregation is
an indication of NARWs?  Regarding the delineation of aggregation in
measurements, an exploration of surface current deformation by SAR
contrast seems well motivated \citep{Kudryavtsev_etal_2012a,
  Rascle_etal_2014}.  However, observations and models predict an
inverse dependence of SAR contrast on wind forcing and sea state, at
least for wind speeds within a range of 1-10~ms$^{-1}$
\citep{Espedal_etal_1998}.  In the Shediac Valley, for example, where
NARWs have been sighted more frequently since 2015, SAR contrast was
found to be weak and wind speed strong relative to other GSL regions.

There has been generally improving SAR coverage of the western GSL
during the 2010s, with Radarsat-2 capturing about a third of the whale
sightings in 2019.  A focus on scenes from mid-May to mid-August
(2008-2020) permitted a characterization of the SAR-wind relationship
when winds are lighter and {\it Calanus} would have been
intermittently aggregated near the surface.  Our definition of SAR
measurement contrast followed \citet{Kudryavtsev_etal_2012b}, but
employed agreement among three overlapping resolution brackets to
isolate contiguous contrasts.  Those that spanned a distance of at
least 10~km (approaching the eddy length scale;
\citealt{Munk_etal_2000}), contributed to an average value of contrast
magnitude on the overlap with the Anticosti, Gasp\'{e}, or Shediac
domains.

Measurement modelling was introduced as a complement to radar imaging
models and ecosystem models that are consistent with a broader set of
measurements, or that provide a general expression of SAR contrast at
the air-sea interface.  This study focused on the strength of the
relationship between SAR contrast magnitude and ERA5 wind speed to
some power, using Pearson and distance correlation
\citep{Szekely_etal_2007, Szekely_Rizzo_2009}.  Because distance
correlation is sensitive to nonlinear and nonmonotonic dependence, it
is more consistent with general expressions of SAR contrast (e.g.,
including surfactants, sea state, and wind direction).  However, both
nonlinear association and a lack of association were also considered,
along with linear and nonlinear components of Pearson correlation.

Correlation magnitude was greatest for wind speed to a positive power,
and Pearson and distance correlation were consistent.  Both were close
to the dominant linear dependence, but were shifted slightly toward a
secondary nonlinear dependence that was apparent by the nonlinear
contribution to Pearson correlation, and by a relatively large value
of distance correlation.  This provided a simple weighting of SAR
contrast whose impact was a correlation of less than 0.2 for all GSL
subdomains.  By making contrast magnitude more uniform (or less
dependent on wind speed), exploration of NARW prey aggregation in a
collection of SAR scenes, (and perhaps sun glint, if IR/color images
are available) may thus be enhanced.  The need to define typical SAR
signatures of biological aggregation in the GSL should also be
informed by complementary measures of contrast in mass and motion
(e.g.~\citealt{Basedow_etal_2019, Barthelmess_etal_2021}), as well as
by observed and simulated measures of {\it Calanus} spp. availability
and supply \citep{Gavrilchuk_etal_2020, Brennan_etal_2021}.

\section*{Acknowledgements}

Comments from Hilary Moors-Murphy, Catherine Brennan, Kevin Sorochan,
three anonymous reviewers, and the associate editor contributed much
to our presentation.  We thank the North Atlantic Right Whale
Consortium and Team Whale of Fisheries and Oceans Canada for their
efforts in collecting and quality controlling the GSL whale sightings.
Radarsat data were produced by MacDonald, Dettwiler and Associates
Ltd.\ and obtained from the Earth Observation Data Management System
of Natural Resources Canada.  Sentinel-1 data were obtained from the
Copernicus programme of the European Union and the European Space
Agency.  The ERA5 data were obtained from the Copernicus Climate
Change Service (C3S) Climate Data Store.  We thank M.~Rizzo and
G.~Sz{\'e}kely for providing the distance correlation R package
(energy).  Funding was provided by the Competitive Science Research
Fund (CSRF) and Species at Risk (also denoted SAR) Program of
Fisheries and Oceans Canada.

\end{multicols}

\end{document}